\pdfoutput=1
\documentclass[12pt,letterpaper]{iopart}
\usepackage[colorlinks=true,linkcolor=blue,urlcolor=blue,citecolor=blue,linktocpage=true]{hyperref}
\usepackage{amssymb}

\makeatletter
\def\@fnsymbol#1{\ifcase#1 \or ^{1} \or ^{2} \or ^{3} \or ^{4} \or %
  ^{5} \or ^{6} \or ^{7} \or ^{8} \or ^{9} \or ^{10} \or ^{11} \or %
  ^{12} \or ^{13} \or ^{14} \or ^{15} \or ^{16} \or ^{17} \or ^{18} %
  \else\@ctrerr\fi\relax}
\makeatother

\usepackage{tocloft}

\def\d{\partial}
\def\sech{{\rm \hspace{1.5pt} sech}}
\def\eq#1 { \begin{equation} #1 \end{equation} }
\def\eqn#1{ \begin{eqnarray} #1 \end{eqnarray} }
\def\nn { \nonumber }
\newcommand{\com}[2]{{\left[ #1 ,\, #2 \right]}}
\def\ket#1{\left|#1\right\rangle}

\begin{document}
%

%
%

\title[Hidden symmetries for transparent de Sitter]
{Hidden symmetries for transparent de Sitter space}

\author{Garrett Compton$^{1}$ and Ian A.~Morrison$^{1}$}

\address{$^1$Department of Physics \& Engineering \\ 
  West Chester University \\
  West Chester, PA, 19383, USA}

\eads{\href{mailto:GC902534@wcupa.edu}{GC902534@wcupa.edu} and
  \href{mailto:imorrison@wcupa.edu}{imorrison@wcupa.edu}}

\begin{abstract}
  It is known that odd-dimensional de Sitter space acts as
  a transparent potential for free fields.
  Previous studies have explained this phenomena by relating
  de Sitter free field equations of motion to the time-independent 
  Schrodinger equation with known transparent potentials.
  In this work we show that de Sitter's transparency is a consequence
  of an infinite set of `hidden' symmetries. These symmetries
  arise from an accidental symmetry for the
  zero-mode of matter fields, as well as the boost isometry of
  de Sitter space.
  For simplicity, we consider the case of massive Klein-Gordon theory.
  We show that the Noether charges associated with these hidden symmetries
  distinguish the two linearly-independent solutions of the
  free field wave equation in the asymptotic past and future of de Sitter.
  Conservation of these charges requires that the asymptotic behavior
  of any solution be identical, up to a constant phase,
  in the future and the past, which is the property of transparency.
  In the quantized theory, these charges act trivially on particle
  states belonging to the in/out vacuum Fock space.
  For particle states constructed from other vacua, the action of
  the charges generates particles.
  We comment on how these hidden symmetries may be present in 
  interacting theories.
\end{abstract}

\noindent{\it Keywords\/}: de Sitter space,
QFT in curved spacetime,
integrability


\tableofcontents
\thispagestyle{empty}


%
%
\markboth{Hidden symmetries for transparent de Sitter space \hfill
  {\normalfont G~Compton and I~A~Morrison}~}{}

\section{Introduction}

%
%

Scattering is a ubiquitous feature of quantum systems.
Due to particle/wave duality, quanta behave as waves when
encountering a barrier, and thus in general both transmit and reflect.
In the context of cosmological inflation,
spacetime geometry serves as a time-dependent barrier for
quantum fluctuations.
Scattering in this context results in cosmological
particle production \cite{Birrell:1982ix}.  
This particle production is observed, indirectly, in measurements
of the cosmic microwave background \cite{Mukhanov:2005sc}.
Thus, scattering plays a fundamental fundamental role in
our understanding of inflation. The ubiquity of particle
production in curved spacetimes poses important challenges
to the formulation of scattering matrix theory for
cosmology (see, e.g., \cite{Witten:2001kn,Marolf:2012kh}). 

%
%

It is therefore quite remarkable that de Sitter spacetime, the maximally
symmetric model of an inflating universe, can behave as
a \emph{transparent} barrier for matter fluctuations.
As described in \cite{Lagogiannis:2011st}, following earlier
work in \cite{Bousso:2001mw}, odd-dimensional global de Sitter space
serves as a transparent barrier for linearized matter fluctuations.
This means that fluctuations of matter fields which are pure
positive frequency 
in the asymptotic past emerge in the asymptotic future
also pure positive frequency. This surprising behavior
suggests that linearized field theories in odd-dimensional
de Sitter contain additional, `hidden' structure which constrains
the theory and prevents scattering on cosmological scales.

%
%

Indeed, \cite{Lagogiannis:2011st} found
considerable additional structure.
This work recognized that the time-dependent part of
linearized equations of motion take the
form of the Schr\"odinger equation with transparent potentials.
For example,
for scalar field fluctuations, the time-dependent part of the
wave equation is equivalent to the  Schr\"odinger equation with a
P\"oschl-Teller potential \cite{Poschl:1933zz}.
This potential
arises in many settings,
including supersymmetric quantum mechanics
\cite{Cooper:1994eh,Gangopadhyaya:2011wka},
the dynamics of instantons in 2-dimensional integrable
models \cite{Goldstone:1974gf,Dashen:1975xh},
the inverse scattering approach to the KdV equation
\cite{Grant:1993mi,Dunajski:2010zz,Babelon:2003qtg}),
and tachyon condensation in gauge theories
\cite{Zwiebach:2000dk,Minahan:2000ff}.
For certain values of its parameters,
the P\"oschl-Teller potential is transparent for all incident waves
(for a modern analysis, see, e.g., \cite{Barut:1987am}).
Reference \cite{Lagogiannis:2011st} showed that the Bogoliubov
coefficients which relate fluctuations
in the asymptotic past and future of de Sitter may in turn be related
to the transmission and reflection coefficients of P\"oschl-Teller
theory.
This provides an explanation for why odd-dimensional de Sitter
space acts as a transparent barrier for linearized fields.


%
%

In this work we further explore structures which
impose transparency on odd-dimensional de Sitter space.
For simplicity, we consider the case of massive
Klein-Gordon theory on a fixed de Sitter background.
We show that, when the spacetime dimension
is odd, this theory has an infinite set of `hidden' symmetries.
To our knowledge, these symmetries have not previously appeared
in the literature. The Noether charges corresponding to these
symmetries distinguish between positive- and negative-frequency
solutions to the equation of motion.
The conservation of these charges requires that solutions have the same
asymptotic behavior, up to a constant phase,
in both the past and future asymptotic regions of de Sitter space.
Thus, the conservation
of these charges enforces transparency.
Our investigation serves as an example of how heightened symmetry
in models of the early universe can result in simple asymptotic
behavior of matter fields -- for another example, see, e.g.,
\cite{Costa:2015bps}.

%
%

Key ingredients in our study have appeared in the literature
before. Mode solution the Klein-Gordon
equation are related to one another via what are known as
{Darboux-Crum transformations} \cite{Matveev:1991aa}.
These transformations are familiar -- in action, if not by name -- from
supersymmetric quantum mechanics.
In the context of de Sitter field theory, these transformations
have been pointed out previously, and in particular, are an
important ingredient in the analysis of \cite{Lagogiannis:2011st}.
We show that these transformations can be regarded as
a consequence of the boost isometry of de Sitter space.
Another key ingredient in our analysis is the presence of an
accidental time-translation symmetry for the zero angular momentum
mode solution to the Klein-Gordon equation in three spacetime dimensions.
This too has been noted previously. However, we show that by
combining Darboux-Crum transformations with this charge, one may
generate an infinite family
of conserved quantities which constrain every de Sitter Klein-Gordon mode,
and which are present in all odd dimensions.

%
%

We also investigate our family of charges
in the context of quantized Klein-Gordon theory.
In particular, we
examine the action of the charges in the Fock spaces built
atop the de Sitter-invariant Mottola-Allen vacua, or ``$\alpha$-vacua''
\cite{Mottola:1984ar,Allen:1985ux}.
In this setting, the statement that de
Sitter is transparent may be phrased as saying that 
the natural ``in'' vacuum (whose mode functions are pure positive
frequency in the asymptotic past) is equivalent to the ``out''
vacuum (whose mode functions are pure positive frequency in the
asymptotic future). Thus, in odd dimensions there is just one
``in/out'' vacuum.
While the existence of the conserved quantities is independent of
the choice of vacuum,
different vacua provide valuable interpretation for our family
of charges.
We show that the family of charges act simply in the Fock space of the
``in/out'' vacuum. Particle states in this basis have wavefunctions
which are pure positive frequency in the asymptotic past and
future, and thus are eigenfunctions of the charges.
In the Fock spaces built from other MA vacua, the wavefunctions
are not eigenfunctions of the charges, and thus
the action of the charges generates particles.

%
%

This paper is organized as follows. We begin in \S\ref{sec:KG}
by reviewing relevant aspects of Klein-Gordon theory in de Sitter
space, including the property of transparency in odd dimensions.
In \S\ref{sec:DC} we describe the Darboux-Crum transformations which
relate different mode solutions of the Klein-Gordon wave equation.
We show that these transformations are related to the boost isometry
of de Sitter space. We also show how
these Darboux-Crum transformations, in the
context of P\"oschl-Teller theory, give rise to an infinite family of
conserved charges which in turn imply transparency.
In \S\ref{sec:symmetries} we derive, via the standard Noether procedure
of classical field theory,
an infinite class of hidden symmetries for Klein-Gordon theory
in de Sitter, as well as their associated charges. We show that the
presence of these charges implies transparency in the de Sitter
context.
Then in \S\ref{sec:quantum} we consider the quantized Klein-Gordon
theory. We construct explicit expressions for the hidden charges,
and we analyze their action on different vacua.
We conclude with a discussion in \S\ref{sec:disc}.

\section{Klein-Gordon fields in de Sitter}\label{sec:KG}

In this section we briefly review relevant aspects of Klein-Gordon
theory in dS space, as well as establish our conventions.
For further introduction, see, e.g.,
\cite{Birrell:1982ix,Mottola:1984ar,Allen:1985ux,Spradlin:2001pw}.

%
%
We consider $(D = d+1)$-dimensional de Sitter spacetime whose line
element in global coordinates takes the form
\eq{ \label{eq:dmetric}
  \frac{ds^2}{\ell^2} = -dt^2 + \cosh^2 t \,d\Omega_d^2 
  = -dt^2 
  + \cosh^2 t \left( d\theta^2 + \sin^2\theta \, d\Omega_{d-1}^2\right) .
}
Here $\ell$ is the de Sitter radius, $t\in\mathbb{R}$ is a
dimensionless global time coordinate,
$d\Omega_d$ is the line element on $S^d$, and
$\theta \in [0,\pi)$ is the polar angle on $S^d$.
As is evident from the line element, the topology of global de Sitter
is $\mathbb{R}\times S^d$. The isometries of de Sitter space correspond
to rotations and boosts.

%
%
We study a real scalar field $\Phi(x)$ satisfying the Klein-Gordon
equation 
\eq{ \label{eq:KGeqn}
  (\Box - M^2) \Phi(x) = 0 ,
}
where $\Box$ is the d'Alembertian operator on de Sitter space
and $M$ is a positive mass satisfying
\eq{ \label{eq:mass}
  M^2 \ell^2 > \frac{d^2}{4} .
}
For our purposes it is
more convenient to consider the rescaled field $\Psi(x)$
related to $\Phi(x)$ via
\eq{ \label{eq:redefinition}
  \Phi(x) = \ell^{(2-D)/2} (\cosh t)^{-d/2} \Psi(x) .
}
We have included factors of the de Sitter radius so as to
make $\Psi(x)$ dimensionless.
We may expand $\Psi(x)$ in a basis of spherical
harmonics on $S^d$,
\eq{ \label{eq:PsiModes}
  \Psi(x) = 
  \sum_{\vec{L}} \Psi_{\vec{L}}(t) Y_{\vec{L}}(\Omega) .
}
Here we denote the $d$ coordinates parameterizing the $S^d$
by the collective coordinate $\Omega$. The harmonics
are labeled by a set of $d$ angular momenta $\vec{L}$ with
the total angular momentum denoted by $L$;
they satisfy the eigenvalue and orthonormality conditions
\eq{
  \triangle_d Y_{\vec{L}}(\Omega) = - L(L+d-1) Y_{\vec{L}}(\Omega) ,
  \quad
  \int d\Omega_d \, Y_{\vec{L}}(\Omega) Y^*_{\vec{L}'}(\Omega)
  = \delta_{\vec{L}\vec{L}'} ,
}
where $\triangle_d$ is the Laplacian on $S^d$ and $\delta_{\vec{L}\vec{L}'}$
is the Kronecker delta symbol. Our conventions for spherical harmonics
are standard and correspond to those of, e.g., \cite{Higuchi:1986wu}.
Upon inserting the mode expansion (\ref{eq:PsiModes}) into
(\ref{eq:KGeqn}), one may obtain the equation of motion for the
time-dependent fields $\Psi_{\vec{L}}(t)$. This equation
depends only upon the dimension $d$, mass $M$, and the
total angular momentum $L$. It is convenient to keep track of 
these parameters with the dimensionless quantities
\eq{ \label{eq:nuAndK}
  \nu := L + \frac{d}{2} - 1, \quad \textrm{and} \quad
  k := \sqrt{M^2\ell^2 - \frac{d^2}{4}} .
}
We refer to $\nu$ as the \emph{level}.
Given our restriction to
sufficiently massive fields (\ref{eq:mass}) it follows that $k>0$.
For simplicity, when there is no risk of confusion
we will simply write $\Psi_{\vec{L}}(t)$ as $\Psi_{\nu}(t)$.
Then the equation of motion for $\Psi_{\nu}(t)$ is 
\eq{ \label{eq:PsiEOM}
  \left[ - \d_t^2 - \nu(\nu+1) \sech^2 t - k^2\right] \Psi_\nu(t) = 0 .
}
We also note for later that this equation has the form of the
time-independent Schrodinger equation (with $t$ playing the role of
position) \cite{Lagogiannis:2011st}.

%
%
A convenient form for the solutions to (\ref{eq:PsiEOM}) is
\eqn{ \label{eq:psiNuK}
  \psi_{\nu k}(t)
  &=& \frac{1}{\sqrt{2 |k|}} e^{- i k t}
  \,{}_2F_1\hspace{-4pt}\left[-\nu,\, \nu+1;\,1+ik;\,\frac{1-\tanh t}{2}\right]   \nn \\
  &=& \frac{1}{\sqrt{2 |k|}} \Gamma(1+ik) (-1)^{i k / 2} P^{-i k}_\nu(\tanh t) .  
}
Here ${}_2F_1[a,b;c;z]$ is the Gauss hypergeometric series,
$\Gamma(x)$ is the Gamma function, and
$P_\mu^\nu(z)$ is the associated Legendre function \cite{abramowitz:1972}.
We note that the expressions (\ref{eq:psiNuK}), like the
equation of motion (\ref{eq:PsiEOM}), are invariant under the replacement
$\nu \to -\nu-1$.
For the values of $\nu$ and $k$ we consider,
the complex conjugate $\psi^*_{\nu k}(t) = \psi_{\nu,-k}(t)$
provides a second linearly independent solution.
These solutions have been normalized so that the
Wronskian satisfies
\eq{ \label{eq:psiWronskian}
  -i \left[\psi_{\nu k}(t) \d_t \psi^*_{\nu k}(t)
    - \psi^*_{\nu k}(t) \d_t \psi_{\nu k}(t)
  \right]_{t = {\rm const.}} = 1 .
}
The fact that the Wronskian is conserved follows from the
the equation of motion. When written in terms of the
Klein-Gordon field $\Phi(x)$, this conserved quantity is
known as the Klein-Gordon flux.

%
%
The behavior of the solutions in the asymptotic regions of
de Sitter is crucial to our study.
In the asymptotic future, the solutions behave as
\eq{ \label{eq:psiFuture}
  \psi_{\nu k}(t \to +\infty)
  =  \frac{1}{\sqrt{2 |k|}} e^{- i k t}\left(1 + O\left(e^{-2|t|}\right)\right),
}
while in the asymptotic past, 
\eqn{ \label{eq:psiPast}
  \psi_{\nu k}(t \to -\infty) &=&
  \frac{1}{\sqrt{2 |k|}}
  \frac{\Gamma(1+i k)\Gamma(i k)}{\Gamma(1+ik+\nu)\Gamma(ik-\nu)}
  e^{- i k t}\left(1 + O\left(e^{-2|t|}\right)\right)
  \nn \\ & &
  +  \frac{1}{\sqrt{2 |k|}}
  \frac{\Gamma(1+i k)\Gamma(-i k)}{\Gamma(1+\nu)\Gamma(-\nu)}
  e^{+ i k t}\left(1 + O\left(e^{-2|t|}\right)\right)   .
}
In particular, we note that $\psi_{\nu k}(t)$ behaves as the
plane wave $e^{-ikt}$ in the asymptotic future. In the asymptotic
past, for generic values of $k$ and $\nu$, $\psi_{\nu k}(t)$ behaves
like a linear combination of the two plane waves $e^{\pm i kt}$.

%
%
Given our basis of solutions, 
it is straightforward to see that the de Sitter background acts
as a transparent potential
when $\nu = n$, where $n\in\mathbb{Z}$.
These values for $\nu$ correspond to when the number of spatial dimensions 
$d$ is even, i.e., when the spacetime dimension is odd.
For $\nu = n$ the hypergeometric series in (\ref{eq:psiNuK})
terminates and is thus an $n$-th
order polynomial in $\tanh t$. Correspondingly, when
$\nu = n$ the second term in (\ref{eq:psiPast}) vanishes.
Thus,
$\psi_{\nu k}(t)$ has the same behavior (up to a constant phase)
in both asymptotic regions.
This is the property of transparency:~a pure positive-frequency
solution in the far past travels through de Sitter
and emerges pure positive frequency.

\section{Darboux-Crum transformations}\label{sec:DC}

%
%
In the previous section, the transparency of odd-dimensional
de Sitter space arose in a rather matter-of-fact manner from
examining solutions to the Klein-Gordon equation.
It turns out that there is much more structure behind this phenomena.
In this section we describe the existence of Darboux-Crum
transformations which relate the modes $\Psi_\nu(t)$
at different levels, i.e., different values of $\nu$.
We show that this Darboux-Crum structure is intimately
related to the boost isometry of de Sitter space.
Then, following \cite{Lagogiannis:2011st}, we describe the relation
between dS Klein-Gordon field theory and supersymmetric
quantum mechanics with P\"oschl-Teller potentials.
This is an ideal setting to demonstrate how
the Darboux-Crum structure can, when combined with a single conserved
quantity, result in an infinite family of conserved quantities
and transparency.

\subsection{Darboux-Crum transformations}

%
%
Recall that the equation of motion (\ref{eq:PsiEOM}) for
$\psi_{\nu k}(t)$ has the form of the time-independent Schr\"odinger
equation. Indeed, we may write this equation as
\eq{ \label{eq:Schrodinger}
  H_\nu \psi_{\nu k}(t) 
  = k^2 \psi_{\nu k}(t) ,
}
where $H_\nu$ is the differential operator
\eq{
  H_\nu 
  := -\d_t^2 - \nu(\nu+1) \sech^2\,t .
}
In analogy with the Schr\"odinger equation, we refer to $H_\nu$
as the Hamiltonian at level $\nu$. In this analogy,
$t$ plays the role of position and $k^2$ plays the role of
energy.
%
%
The Hamiltonian may usefully be written in terms of
the differential operators
\eqn{ \label{eq:Apm}
  A^+_\nu &=& - \d_t + (\nu+1)\tanh t, \\
  A^-_\nu &=& \d_t + \nu \tanh t .
}
In terms of these operators, $H_\nu$ may
be written variously as
\eq{
  H_\nu = A^-_{\nu+1} A^+_\nu - (\nu+1)^2 = A^+_{\nu-1} A^-_\nu - \nu^2 .
}
The operators $A^\pm_\nu$ relate  Hamiltonians whose index
$\nu$ differs by one:
\eq{ \label{eq:intertwining}
  H_{\nu+1} A_\nu^+ = A_\nu^+ H_\nu , \quad
  H_\nu A^-_{\nu+1} = A^-_{\nu+1} H_{\nu+1} .
}
The relations (\ref{eq:intertwining}) are known
as intertwining relations, and we will refer to $A^\pm_\nu$
as intertwining operators.
The intertwining operators act as raising/lowering operators
for solutions in that they change the value of the level $\nu$ by
one, i.e.,
\eq{ \label{eq:AonPsi}
  A^+_\nu \psi_{\nu k}(t) = (ik+\nu+1) \psi_{\nu+1,k}(t) , \quad
  A^-_\nu \psi_{\nu k}(t) = (-ik +\nu) \psi_{\nu-1,k}(t) .
}
These relations may be verified explicitly by, e.g., utilizing the
well-known recursion relations of Legendre functions \cite{abramowitz:1972}.

%
%
Transformations between classes of solutions of the form
(\ref{eq:AonPsi}) are known as {Darboux-Crum transformations}
\cite{Matveev:1991aa}. We note that, due to our restriction to
$\nu\in\mathbb{R}$ and $k>0$, the transformations (\ref{eq:AonPsi})
do not annihilate any wavefunctions. In addition, when mapping
wavefunctions from one level to another, these transformations preserve
the spectrum of energy eigenvalues. 
Thus, these Darboux-Crum transformations provide an isomorphism
between PT theories whose level differs by unity.
In the language of SUSYQM, these transformations are known
as isospectral deformations
\cite{Cooper:1994eh,Gangopadhyaya:2011wka}.\footnote{In the context
  of supersymmetric quantum mechanics, it is natural to consider all
  values of $k$ which result in a normalizable wavefunction. In this
  case, the Darboux-Crum transformations we describe can introduce or remove
  bound states, i.e. states with $k^2<0$. Thus, in this more
  general context, the Darboux-Crum transformations are referred to
  as quasi-isospectral deformations.}  

\subsection{Boost symmetry}

Before proceeding to analyze the consequences of the
Darboux-Crum transformations just described, we pause here to
show that these transformations may be viewed as a consequence
of the boost isometry of de Sitter space.

%
%
Recall that the isometries of de Sitter space may be described
as rotations and boosts. For our global coordinates (\ref{eq:dmetric}),
the former act on the $S^d$ coordinates while preserving the
time foliation; the latter
involve both time and spatial coordinates thus alter the time foliation.
One such boost Killing vector is
\eq{ \label{eq:xi}
  \xi^\mu \d_\mu = \cos\theta \, \d_t - \tanh t \sin \theta \, \d_\theta .
}
Under an infinitesimal boost along $\xi^\mu$,
the Klein-Gordon field $\Phi(x)$ transforms as
\eq{
  \Phi(x) \to \Phi(x) + \epsilon \mathcal{L}_\xi \Phi(x),
}
where $|\epsilon| \ll 1$ is an infinitesimal constant and
$\mathcal{L}_\xi$ denotes the Lie derivative along $\xi^\mu$.
Since the boost is an isometry, the transformed field is also
a solution to the Klein-Gordon equation; in particular,
$\mathcal{L}_\xi \Phi(x)$ is itself a solution. Thus, this infinitesimal
boost provides a map between solutions to the Klein-Gordon equation.

The rescaled field $\Psi(x)$ transforms under this boost as
\eq{ \label{eq:PsiBoost}
  \Psi(x) \to \Psi(x) + \epsilon
  \left(-\frac{d}{2} \tanh t \cos\theta + \mathcal{L}_\xi\right)\Psi(x) .
}
Let us examine how this transformation effects a single
mode solution 
$\Psi_{\vec{L}}(t) Y_{\vec{L}}(\Omega)$.
Since the boost does not preserve the
global time foliation, it mixes the
modes of $\Psi(x)$.
In particular, the transformation (\ref{eq:PsiBoost}) takes a
single mode into two terms, one with a total angular momentum
raised by one, and one with a total angular momentum lowered by one. 
Explicitly, denoting the angular momenta
as $\vec{L} = (L,\vec{m})$, the second term in 
(\ref{eq:PsiBoost}) is
\eqn{ \label{eq:BostOfPhi}
  \delta \left(\Psi_{L\vec{m}}(t) Y_{L\vec{m}}(\Omega)\right)
  &:=& \left(-\frac{d}{2} \tanh t \cos\theta + \mathcal{L}_\xi\right)
  \left(\Psi_{L \vec{m}}(t) Y_{L \vec{m}}(\Omega)\right) \nn \\
  &=& b^-_{Lm} \Psi_{L-1, \vec{m}}(t) Y_{L-1, \vec{m}}(\Omega) 
- b^+_{Lm} \Psi_{L+1, \vec{m}}(t) Y_{L+1, \vec{m}}(\Omega) , \nn \\
}
where the coefficients are reported in \cite{Marolf:2008hg} to
be
\eqn{
  b^+_{Lm} &=& 
  \left[ i k + \left(L+\frac{d}{2}\right)\right]
  \left[\frac{(L+m+d-1)(L-m+1)}{(2L+d-1)(2L+d+1)}
  \right]^{1/2},  \\
  b^-_{Lm} &=&
  \left[ -i k + \left(L-1+\frac{d}{2}\right)\right]
  \left[\frac{(L+m+d-2)(L-m)}{(2L+d-1)(2L+d-3)}
  \right]^{1/2} , 
}
where $m = |\vec{m}|$.
Using the orthogonality of the spherical harmonics it is possible to
to isolate the element of the boost which acts on $\Psi_L(t)$ to
raise/lower the value of $L$. For instance,
\eqn{
  & &
  - \left[\frac{(2L+d-1)(2L+d+1)}{(L+m+d-1)(L-m+1)}
  \right]^{1/2}
  \int d\Omega\, Y^*_{L+1,\vec{m}}(\Omega)
  \delta \left(\Psi_{L \vec{m}}(t) Y_{L \vec{m}}(\Omega)\right)
  \nn \\
  & & \quad
  = \left(- \d_t + \left(L + \frac{d}{2}\right)\tanh t\right)
  \Psi_{L \vec{m}}(t) = \Psi_{L+1,\vec{m}}(t) . 
}
We recognize the differential operator after the first equality
to be $A_\nu^+$ (written in terms of $L$ and $d$). In a similar
way, one obtains $A_\nu^-$ by isolating the part of the boost
which lowers $L$ by one.

To summarize, the intertwining operators $A_\nu^\pm$ agree with the
action of an infinitesimal boost isometry on the time-dependent
part of the mode solutions for $\Psi(x)$. Thus, we regard the
Darboux-Crum transformations which map mode solutions at different
levels as a consequence of the
boost isometry of de Sitter space.
We note that the Darboux-Crum structure exists for scalar fields
in all spacetime dimensions,
and also for other linearized fields on de Sitter,
including spin-half, symmetric tensor, and p-form fields
\cite{Lagogiannis:2011st}.
In all these cases, the Darboux-Crum structure may be regarded
as a consequence the boost isometry combined with the linear nature of
the equation of motion.

\subsection{Charges in P\"oschl-Teller theory}

%
%
We now return to our discussion of the Sch\"odinger equation
(\ref{eq:Schrodinger}) 
which provides the equation of motion for the mode solutions
$\psi_{\nu k}(t)$. The potential which appears in this equation,
\eq{ \label{eq:Vnu}
  V_\nu(t) = - \nu(\nu+1)\sech^2 t,
}
is known as the P\"oschl-Teller (PT) potential
\cite{Poschl:1933zz,Barut:1987am}. We refer to the family of quantum
mechanical theories labeled by $\nu$ as the PT family.
The mode solutions $\psi_{\nu k}(t)$ provide the wavefunctions
corresponding to scattering states for this potential. 
The intertwining operators $A^\pm_\nu$ map the Hamiltonian
and wavefunctions of the theory at level $\nu$ to those of the theory
with level $\nu\pm 1$.
When $\nu = n$, where $n \in \mathbb{Z}$, the
potentials (\ref{eq:Vnu}) are transparent.
A simply way to determine this is by calculating the transmission
and reflection coefficients of the theory
\cite{Cooper:1994eh,Gangopadhyaya:2011wka}.
For our purposes, it is more enlightening to examine the role
that conserved quantities play in enforcing transparency.

%
%
Suppose that at level $\nu$ (not necessarily an integer) there exists
a conserved quantity,
i.e., an operator $Q_\nu$ which commutes with the level $\nu$ Hamiltonian,
\eq{
  \com{Q_\nu}{H_\nu} = 0 .
}
Using the intertwining operators $A^\pm_\nu$ we may define a conserved
quantity in the $\nu+1$ theory as
\eq{ \label{eq:Qnplus1}
  Q_{\nu+1} := A^+_\nu Q_\nu A^-_{\nu+1} .
}
We check explicitly that this commutes with the level $\nu+1$
Hamiltonian:
\eqn{
  \com{Q_{\nu+1}}{H_{\nu+1}}
  &=& Q_{\nu+1} H_{\nu+1} - H_{\nu+1} Q_{\nu+1} \nn \\
  &=& A_\nu^+ Q_{\nu} A^-_{\nu+1} H_{\nu+1} - H_{\nu+1} A^+_\nu Q_\nu A^-_{\nu+1} \nn \\
  &=& A_\nu^+ \com{Q_\nu}{H_\nu} A_{\nu+1}^- \nn \\
  &=& 0 .
}
The second equality follows from the definition of $Q_{\nu+1}$;
the third equality follows from the intertwining relations;
and the forth equality follows from the fact that $Q_{\nu}$ is conserved
at level $\nu$.
In a similar manner, we can likewise define a conserved quantity
at level $\nu - 1$ via
\eq{ \label{eq:Qnminus1}
  Q_{\nu-1} := A_{\nu}^- Q_\nu A^+_{\nu-1} .
}
It follows that through repeated application of (\ref{eq:Qnplus1})
or (\ref{eq:Qnminus1}) we can construct a conserved quantity
at all levels $\nu + j$, $j \in \mathbb{Z}$.\footnote{We remind
  the reader that the PT potential (\ref{eq:Vnu}) is invariant
  under the relabeling $\nu \to -\nu - 1$, and so the levels
$\nu$ and $-\nu-1$ define equivalent theories.} Thus, the
existence of a single
$Q_\nu$ implies an infinite family of conserved quantities, one in
each level $\nu+j$. We emphasize that this structure occurs
for all real $\nu$.

%
%
The concrete family of conserved quantities relevant to our study occurs
for $\nu = n$.
We start in the $n = 0$ theory,
where the potential $V_0(t) = 0$ is trivial. In this theory
the linear momentum operator, which we denote
\eq{
  Q_0 = i \d_t ,
}
is a conserved quantity. The wavefunctions of this theory
are the left- and right-moving plane waves $e^{\pm i k t}$.
The eigenvalues of $Q_0$ are $\pm k$, and so eigenfunctions of $Q_0$
must exhibit the same left- or right-moving behavior everywhere.
In this way, one may say that the $n=0$ theory is transparent
as a consequence of the existence of the conserved quantity $Q_0$.

At higher levels $n>0$ the PT potential is non-trivial and
the theories do not enjoy conserved linear momentum.
However, from $Q_0$ we may construct a conserved quantity at
each level $n$, simply
by repeatedly applying the procedure (\ref{eq:Qnplus1}).
The result is a conserved quantity $Q_n$ at each level $n$:
\eq{
  Q_n := A_{n-1}^+ \dots A_0^+
  Q_0
  A_1^- \dots A_n^- .
}
The action of $Q_n$ on the wave functions can be determined
by using the raising and lowering relations (\ref{eq:AonPsi});
the result is
\eq{
  Q_n \psi_{n k}(t) = k \left[\prod_{j=1}^n(k^2+j^2)\right] \psi_{nk}(t) .
}
We see that $Q_n$, like $Q_0$, distinguishes between $\pm k$
eigenvalues. Let us see how this affects the asymptotic behavior
of the wavefunctions.
For $n>0$ the wave functions are not simple plane waves. Nevertheless,
at asymptotically large values of $|t|\gg 1$, the potential is
exponentially suppressed, and the wave functions are 
plane waves up to exponentially suppressed corrections, i.e.,
\eq{ \label{eq:asymptoticQM}
  \psi_{n k}(|t|\gg 1) \propto e^{\pm i k t}\left(1 + O(e^{-2|t|})\right) .
}
From this asymptotic behavior we see that
the eigenfunctions of $Q_n$ must exhibit the
same plane wave behavior in both regions $t \to - \infty$ and
$t \to +\infty$. Thus, the presence of $Q_n$ enforces 
transparency for the non-trivial potential $V_n(t)$. 
Since there exists a $Q_n$ at each level $n \in \mathbb{Z}$, it follows
that all PT theories with level $n \in \mathbb{Z}$ are transparent.

\section{Hidden symmetries in dS}\label{sec:symmetries}

%
%
We now return to our discussion of Klein-Gordon
theory on de Sitter.
We focus on the case when the spacetime dimension is odd,
so we let $\nu = n$ with $n \in \mathbb{Z}$.
We will find that there exist an infinite family of
`hidden' Noether symmetries. The conserved quantities corresponding
to these symmetries are analogues of the $Q_n$ charges
of PT theory described in the previous section. 
However, we emphasize that our analysis in this section is purely
classical. 

%
%
The action for the Klein-Gordon field $\Phi(x)$ is
\eq{ \label{eq:SKG}
  S[\Phi] = - \frac{1}{2}\int d^Dx\sqrt{-g(x)}
  \left( g^{\mu\nu}(x)\d_\mu \Phi(x) \d_\nu \Phi(x)
  + \frac{M^2}{2} \Phi^2(x) \right) .
}
To bring this into a more useful form, we first replace
$\Phi(x)$ for $\Psi(x)$ as in (\ref{eq:redefinition}),
insert the mode expansion for $\Psi(x)$ (\ref{eq:PsiModes}),
then integrate over the $S^d$. After these steps the
action becomes a sum of terms, each of which is quadratic in a 
single $\Psi_{\vec{L}}(t)$; schematically, we write this as
\eq{ \label{eq:Ssum}
  S[\Phi] = \sum_{\vec{L}} S_{\vec{L}}[\Psi_{\vec{L}}] .
}
The expression for $S_{\vec{L}}[\Psi_{\vec{L}}]$ can be tidied up
with a bit of algebra, and by dropping a total derivative.
Then the expression for $S_{\vec{L}}[\Psi_{\vec{L}}]$ involves
only $M$, $d$, and $L$, and so is most conveniently written
in terms of $k$ and $n$ as
\eqn{
  S_n[\Psi_n] &=& \frac{1}{2} \int_{-\infty}^\infty dt
  \left[
    (\d_t\Psi_n)^2 - n(n+1) \sech^2 t \,\Psi_n^2 - k^2 \Psi_n^2
  \right] , \nn \\
  &=& \frac{1}{2} \int_{-\infty}^\infty dt \, \Psi_n H_n \Psi_n .
}
In the second line we have identified the same differential
operator $H_n$ which plays the role of the Hamiltonian in the
context of PT theory. We remind the reader that our
short hand $\Psi_n(t)$ denotes any mode
$\Psi_{\vec{L}}(t)$ with total angular momentum $L$.

%
%
We start by examining the $n=0$ case, which corresponds 
to the zero angular momentum mode in $D=3$ dimensions.
The action for $\Psi_0(t)$ is explicitly
\eq{
  S_0 = \frac{1}{2} \int_{-\infty}^\infty dt \left(
    (\d_t\Psi_0)^2 - k^2 \Psi_0^2 \right) .
}
We recognize this as the action for a free harmonic oscillator;
the presence of the de Sitter background has been completely
absorbed by the field redefinition (\ref{eq:redefinition})
exchanging $\Phi(x)$ for $\Psi(x)$.
Since there is no explicit time dependence,
this action enjoys time-translation symmetry.
The infinitesimal transformation
\eq{
  \Psi_0 \to \Psi_0 + \delta\Psi_0, \quad
  \delta\Psi_0 = \epsilon \d_t \Psi_0 , \quad |\epsilon|\ll 1,
}
leaves the action invariant up to a total derivative.
Indeed, it is easy to show that
\eq{ \label{eq:deltaS0}
  \delta S_0
  = \frac{1}{2} \epsilon \int_{-\infty}^\infty dt \, \d_t
  \left((\d_t\Psi_0)^2 - k^2 \Psi_0^2\right) 
}
follows simply from algebraic manipulation
(that is, $\Psi_0$ need not satisfy any equation of motion).
It is remarkable that
a degree of freedom enjoys time-translation symmetry
on a de Sitter background, which is itself not invariant under
time translations. 

%
%
Through successive Darboux-Crum transformations it is possible
to construct a Noether symmetry at each level $n$, built from
the $n=0$ time translation symmetry.
For the mode $\Psi_n(t)$, consider the variation
\eq{
  \delta \Psi_n = \epsilon \mathcal{D}_n \Psi_n ,
}
where $\mathcal{D}_n$ is the derivative operator defined recursively
as
\eqn{
  \mathcal{D}_0 &=& \d_t, \\
  \mathcal{D}_n &=& A_{n-1}^+ \mathcal{D}_{n-1} A^-_n
  = A_{n-1}^+\dots A_0^+ \d_t A_1^-\dots A_n^- ,
  \quad n>0 .
}
Then
\eq{
  \delta \Psi_n = \epsilon \mathcal{D}_n \Psi_n
  = \epsilon A_{n-1}^+ \mathcal{D}_{n-1} A_n^- \Psi_n .
}
With this variation of the field, the variation of the
level $n$ action is
\eqn{
  \delta S_n[\Psi_n]
  &=& \int_{-\infty}^\infty dt\, \Psi_n H_n \delta \Psi_n 
  = \epsilon \int_{-\infty}^\infty dt\, \Psi_n H_n
  A_{n-1}^+ \mathcal{D}_{n-1} A_n^- \Psi_n  .
}
We may use the intertwining relation (\ref{eq:intertwining}) to replace
$H_n A^+_{n-1} = A^+_{n-1} H_{n-1}$,
then integrate by parts so that
$ \Psi_n A^+_{n-1}\dots$ may be replaced with $(A^-_n \Psi_n)\dots\,$.
The result is
\eqn{
  \delta S_n[\Psi_n]
  &=& \epsilon \int_{-\infty}^\infty dt
  \left( A_n^- \Psi_n \right)
  H_{n-1} \mathcal{D}_{n-1} A^-_n \Psi_n . 
}
Upon defining a new field
\eq{
  \chi(t) := A_n^- \Psi_n(t) ,
}
then $\delta S_n$ takes the form
\eq{
  \delta S_n[\Psi_n] = \epsilon \int_{-\infty}^\infty dt\,
  \chi H_{n-1} \left( \mathcal{D}_{n-1} \chi \right)
   = \delta S_{n-1}[\chi] .
}
Thus, the variation of the level $n$ action
is equivalent to the variation of the level $n-1$ action. Since
$\delta S_0[\chi]$ is a total derivative, it follows by induction
that for all $n$ the variation $\delta S_n[\chi]$ is also a total
derivative. This proves that the transformations
\eq{ \label{eq:PsiSymm}
  \Psi_n(t) \to \Psi_n(t) + \epsilon \mathcal{D}_n \Psi_n(t)
}
are Noether symmetries.

%
%
The conserved quantities associated with these symmetries may
be constructed by the usual Noether procedure. Although
we are interested in the theory of the real field $\Psi(x)$,
it is convenient to construct expressions for conserved quantities
valid for complex fields as well, so that we may evaluate these
quantities of our basis solutions $\psi_{nk}(t)$ which are complex.
For the $n=0$ symmetry the conserved quantity is
\eq{ \label{eq:P0}
  P_0 := \left|\d_t\Psi_0\right|^2 + k^2 |\Psi_0|^2 ,
}
which we recognize as the usual time translation operator.
For $n>0$, the conserved quantity associated to (\ref{eq:PsiSymm}) is
\eq{ \label{eq:Pn}
  P_n := 
    \left|\d_t\left(A_1^-\dots A_n^- \Psi_n\right)\right|^2
    + k^2 \left| A_1^-\dots A_n^- \Psi_n\right|^2 .
}
In essence, this is the conserved quantity
$P_0$ constructed from a higher-level field $\Psi_n(t)$ by performing
multiple Darboux-Crum transformations to lower the field to level $n=0$.

%
%
We now show that the existence of the conserved quantities
$P_n$ implies the property of transparency.
Our argument is similar to the one we employed to show that
the existence of the conserved quantities $Q_n$ imply
transparency in PT theory.
For $n=0$, the conservation of $P_0$ implies time translation symmetry
which in turn implies that there is no potential, and thus nothing
for $\Psi_0$ to scatter off.
For the $n>0$, we examine
asymptotic solutions to the wave equation. Suppose that
a level $n$ solution $\psi(t)$ is pure positive frequency in
the asymptotic future, i.e.,
\eq{
  \psi(t\to+\infty) = \frac{1}{\sqrt{2|k|}}
    e^{-i k t} + O\left(e^{-2|t|}\right) .
}
In the asymptotic past, the solution could, a priori, contain
both positive and negative frequency branches, 
\eq{
  \psi(t\to-\infty) = \frac{\alpha}{\sqrt{2|k|}} e^{-i k t}
  + \frac{\beta}{\sqrt{2|k|}} e^{+ i k t} + O\left(e^{-2|t|}\right) .
}
Conservation of Klein-Gordon flux, i.e., the Wronskian
(\ref{eq:psiWronskian}),
constrains the coefficients $\alpha$ and $\beta$ to
satisfy
\eq{ \label{eq:fluxCondition}
  |\alpha|^2 - |\beta|^2 = 1 .
}
Evaluating $P_n$ on these asymptotic solutions, one obtains
\eq{
  P_n\left[\psi(t\to+\infty)\right] = |k| \prod_{j=1}^n(k^2 +j^2) ,
}
and
\eq{
  P_n\left[\psi(t\to-\infty)\right] =
  \left(|\alpha|^2 + |\beta|^2\right)
  |k| \prod_{j=1}^n(k^2 +j^2) .
}
From these expressions we see that conservation of $P_n$ requires
\eq{ \label{eq:chargeCondition}
  |\alpha|^2 + |\beta|^2 = 1 .
}
The two requirements (\ref{eq:fluxCondition}) and (\ref{eq:chargeCondition})
are satisfied only for $\beta = 0$ and $|\alpha|^2 = 1$.
Thus, conservation of $P_n$ implies that
any solution at level $n$  which is pure positive frequency in
the future must also be pure positive frequency in the past,
which is the property of transparency. Since exists a conserved
quantity at each level $n$, i.e., there is a conserved quantity
$P_{\vec{L}}$ corresponding to each value of angular momentum $\vec{L}$,
it follows that the de Sitter background is transparent.

\section{Quantization}\label{sec:quantum}

%
%
In this section we turn to quantized Klein-Gordon theory
and examine the family of charges found above in this context.
We will see that the charges act trivially in the Fock space
generated from the in/out vacuum. In other Fock spaces,
the charges create particles, and so no finite-particle
eigenstates of the charges exist in these spaces.

%
%
To quantize the field $\Psi(x)$
we expand in a basis of solutions
\eq{ \label{eq:PsiExpansion}
  \Psi(x)
  = \sum_{\vec{L}} \left[ a_{\vec{L}} \psi_{\vec{L}}(t) Y_{\vec{L}}(\Omega)
    + a^\dagger_{\vec{L}} \psi^*_{\vec{L}}(t) Y^*_{\vec{L}}(\Omega) \right] .
}
In this expression we have restored the angular momenta labels;
$\psi_{\vec{L}}(t)$ are the same solutions $\psi_{nk}(t)$,
i.e.~(\ref{eq:psiNuK}), used throughout our study. 
For a real field, $a_{\vec{L}}$ and $a_{\vec{L}}^\dagger$
are hermitian conjugates.
Upon canonical quantization, the coefficients
$a_{\vec{L}}$, $a_{\vec{L}}^\dagger$ are
promoted to creation and annihilation operators which
satisfy the canonical commutation relations
\eq{
  \com{a_{\vec{L}}}{a^\dagger_{\vec{L}'}}
  = \delta_{\vec{L}\vec{L}'}, \quad
  \com{a_{\vec{L}}}{a_{\vec{L}'}} =
  \com{a^\dagger_{\vec{L}}}{a^\dagger_{\vec{L}'}} = 0 .
}
We define a vacuum state $\ket{\Omega}$ as the state for which
\eq{
  a_{\vec{L}} \ket{\Omega} = 0 , \quad \forall \;\vec{L} .
}
Then the 1-particle Fock space built from $\ket{\Omega}$ is
\eq{
  \mathcal{H}_1 := \left\{
    |\vec{L}\rangle = a_{\vec{L}}^\dagger\ket{\Omega} \; \forall \; \vec{L}
  \right\} .
}

%
%
As is well known, the canonical quantization procedure
described above depends upon the choice of basis solutions
used to expand the field in (\ref{eq:PsiExpansion}).
Different choices for the basis solutions results in different
vacua. The most common choices of vacuum correspond to
the family of de Sitter-invariant vacua known as
Mottola-Allen (MA) vacua (or $\alpha$-vacua)
\cite{Mottola:1984ar,Allen:1985ux}.
In many applications, the most logical vacuum is that of the
Hartle-Hawking\footnote{This state is also known as
  the Bunch-Davies or Euclidean state. This is the state whose
  correlation functions may
  be obtained by analytic continuation from Euclidean signature.
} state which is in
thermal equilibrium with the background geometry
\cite{Gibbons:1977mu}.
In contrast, the modes $\psi_{\vec{L}}(t)$ given in (\ref{eq:psiNuK}),
which are pure positive frequency in the asymptotic future and past,
define the de Sitter-invariant ``in/out'' vacuum
$\ket{\Omega}$.\footnote{While not relevant to
  our discussion, we note that the in/out vacuum has
  features which make it undesirable for many applications.
  Like all MA vacua excepting the Hartle-Hawking state,
  correlation functions of this state have ultraviolet (UV) behavior
  which differs from the Hadamard form \cite{Wald:1995yp},
  and so differs from that of the usual Minkowski vacuum
  \cite{Brunetti:2005pr}.
  The Hadamard form is a necessary ingredient in established
  approaches to axiomatic quantum field theory in curved
  spacetime (see, e.g., \cite{Hollands:2008vx,Hollands:2014eia} and
  references therein).
  The non-Hadamard UV behavior of the in/out vacuum
  causes various subtleties even for linearized fields
  (e.g., \cite{Einhorn:2002nu,deBoer:2004nd}), and presents
  significant challenges to formulating perturbative interactions
  \cite{Einhorn:2003xb,Goldstein:2003ut,Goldstein:2003qf}.
  Nevertheless, MA vacua like the in/out vacuum may play an
  interesting role in approaches to a dS/CFT correspondence
  \cite{Strominger:2001pn,Bousso:2001mw}.
  }


%
%
The creation and annihilation operators of different
vacua are related via Bogoliubov transformations.
For example, if $\overline{a}_{\vec{L}}$, $\overline{a}_{\vec{L}}^\dagger$
denote the creation and annihilation operators of another MA vacuum
$\ket{\overline{\Omega}}$, then
\eq{ \label{eq:BoogyBoogy}
  a_{\vec{L}} = \alpha_L \overline{a}_{\vec{L}}
  +\beta_L \overline{a}^\dagger_{\vec{L}} ,
}
where the Bogoliubov coefficients satisfy
\eq{
  |\alpha_L|^2 - |\beta_L|^2 = 1 .
}
Explicit formulas for these coefficients 
may be found in, e.g., \cite{Bousso:2001mw}. When $\beta_L \neq 0$,
as is the case for different MA vacua, one vacuum will contain
particles relative to another's particle basis. Casually, we may
view the state $\ket{\overline{\Omega}}$ as an infinite-particle
state in the particle basis of $\ket{\Omega}$, and vice
verse. Technically, however, the Bogoliubov transformation which
relates these states may not be implementable as a unitary
transformation, and so two MA vacua may not exist in the
same Fock space \cite{Mottola:1984ar}.

%
%
We may obtain an expression for the $P_{\vec{L}}$ charges in the Fock space
of the in/out vacuum by inserting
the expansion (\ref{eq:PsiExpansion}) into (\ref{eq:Pn}).
This yields
\eq{ \label{eq:PL}
  P_{\vec{L}} = |k| \left[\prod_{j=1}^n(k^2+j^2)\right]
  \left( a^\dagger_{\vec{L}} a_{\vec{L}}
    + a_{\vec{L}} a^\dagger_{\vec{L}} \right) , 
}
where as usual $n = L + d/2 - 1$.
It is natural to normal order this operator with respect to
in/out vacuum. This amounts to subtracting the vacuum expectation
value:
\eq{
  :\hspace{-3pt}P_{\vec{L}}\hspace{-2pt}:_{\Omega}\,
  = P_{\vec{L}} - \langle{\Omega}| P_{\vec{L}}|\Omega\rangle 
  = |k| \left[\prod_{j=1}^n(k^2+j^2)\right] a^\dagger_{\vec{L}}
  a_{\vec{L}} .
}
We see that the normal-ordered charges are proportional to
the number operator $N_{\vec{L}} = a_{\vec{L}}^\dagger a_{\vec{L}}$
which counts the number of quanta with angular momentum $\vec{L}$.
Thus, the $P_{\vec{L}}$ simply on 1-particle states; indeed,
their action is analogous to that of the $Q_n$ operators in
PT theory. The $P_{\vec{L}}$ also have  
trivial co-product on the multi-particle Fock space. 
The simple action of the $P_{\vec{L}}$ in this Fock space can
be attributed to the fact that the basis solutions which define
particle states have only a single asymptotic behavior in the past
and future, and thus define eigenstates of the $P_{\vec{L}}$.

%
%
An expression for the $P_{\vec{L}}$ charges in other Fock spaces
may be obtained by transforming the creation and annihilation operators
in (\ref{eq:PL})
according to (\ref{eq:BoogyBoogy}), then normal ordering with respect
to the new vacuum. The result takes the form
\eqn{
	:\hspace{-3pt}P_{\vec{L}}:\hspace{-2pt}_{\overline{\Omega}}
	&=&  |k| \left[\prod_{j=1}^n(k^2+j^2)\right] \nn \\
        & & \times \left[
	\left(|\alpha_L|^2 + |\beta_L|^2\right) 
	\overline{a}^\dagger_{\vec{L}} \overline{a}_{\vec{L}} 
	+ 2 \alpha_L^* \beta_L
	\overline{a}^\dagger_{\vec{L}} \overline{a}^\dagger_{\vec{L}} 
	+ 2 \alpha_L \beta_L^* \overline{a}_{\vec{L}} \overline{a}_{\vec{L}} 
	\right].
}	
This expression is sufficient to see that in Fock spaces other
than the in/out Fock space, the $P_{\vec{L}}$ charges will generate
particles, even when acting on the vacuum $\ket{\overline{\Omega}}$.
Thus, there are no finite-particle number eigenstates of
$:\hspace{-3pt}P_{\vec{L}}:\hspace{-2pt}_{\overline{\Omega}}$ in
these Fock spaces.

\section{Discussion}\label{sec:disc}

In this paper we have shown that massive Klein-Gordon theory
on an odd-dimensional de Sitter background enjoys an infinite
set of symmetries which are `hidden' in the sense that they
do not generate isometries nor are they internal symmetries of
the field theory. Each symmetry acts on a single Klein-Gordon
mode. Correspondingly, there is a Noether charge $P_{\vec{L}}$
for each value of angular momentum.
%
%
Conservation of these charges requires that the solutions
to the Klein-Gordon equation have the same asymptotic behavior,
up to a phase, in the asymptotic past and future.
Upon quantization, the quantum charges
$:\hspace{-3pt}P_{\vec{L}}\hspace{-3pt}:$ act simply on particle states
belonging to the Fock space of the in/out vacuum (these are eigenstates
of the charges).
In the Fock spaces
of all other Mottola-Allen vacua, the charges generate particles.

%
%
The construction of our family of conserved quantities relies on
two ingredients: first, Darboux-Crum transformations
which relate field modes, and second,
the existence of a conserved quantity $P_{\vec{0}}$ which acts
only on the zero angular momentum mode. We have shown 
that the former may be regarded as a consequence of the boost
isometry of the de Sitter background. Indeed, an alternative
way to construct an infinite family of charges is to    
boost the charge $P_{\vec{0}}$. It is easy to see that boosts
do not commute with $P_{\vec{0}}$. For instance,
if one represents $P_{\vec{0}}$ as a flux integral through a surface
of constant global time, then an infinitesimal boost acting
on this charge deforms the Cauchy surface and defines a new
conserved quantity. By repeatedly boosting $P_{\vec{0}}$, i.e.,
by repeatedly commuting $P_{\vec{0}}$ with a boost generator $B$,
one can construct an infinite family of charges which act
upon all modes of the Klein-Gordon field. 
This family of charges is not identical to the family we have
constructed -- our construction results in less
cumbersome expressions for charges -- but the two families
are quite analogous.
The upshot is that a conserved quantity constructed from the zero
mode alone, when combined with a boost isometry, %
results in an infinite family of conserved quantities.

%
%
We expect quite similar results to hold for other linearized fields
on de Sitter. Indeed, much of the groundwork in understanding these
cases has already be laid by \cite{Lagogiannis:2011st}
which demonstrated
that spin-half, symmetric tensor, and p-form fields
all enjoy the Darboux-Crum structure crucial to our analysis.
These authors also show that these fields are governed by
transparent potentials when the spacetime dimension is odd.
Thus, we expect that all such linearized fields enjoy an infinite
family of conserved charges, at least when the spacetime dimension is odd.
It would be interesting to see these charges in detail.

%
%
We have confined attention to scalar fields with positive mass
such that $k^2 = M^2\ell^2 - d^2/4 > 0$, i.e., with masses greater
than the de Sitter scale.
Such positive mass fields belong to the principle series
of scalar representations of the de Sitter isometry group.
Restricting to real $k$ ensures that wavefunctions
of the PT potentials are scattering states, i.e., states
whose energy is greater than the asymptotic value of the potential.
It would be interesting to consider lighter, yet still positive mass,
fields satisfying $M^2\ell^2 > 0$ but $k^2 = 0$, which belong to the
complementary series of scalar representations of the de Sitter group.
For these fields, the associated wavefunctions in PT theory contain
exponential growth far from the potential, and thus are not typically
regarded as physical wavefunctions in quantum mechanics.
Nevertheless, we expect the charge structure to exist, at least for
$k^2 \neq - \mathbb{Z}$.
When $k^2 \neq - \mathbb{Z}$, the eigenvalues of the charges
$Q_n$ in quantum mechanics (similarly, $P_{\vec{L}}$ in de Sitter)
include zero, and we expect exceptional behavior to occur.
Indeed, these values of $k$ correspond
to \emph{bound states} in PT theory. Bound states
have different asymptotic behavior on either side of the potential
well; thus, for these values of $k$ the potential is not transparent.
Similarly, it would be interesting to examine the case of a
massless field, which corresponds to $k^2 = -d^2/4$.

%
%
We close by commenting on how a similar charge structure might arise
in an interacting field theory on de Sitter.
The key ingredients needed to construct the family of charges
were the boost isometry of de Sitter space and a charge which
acts on the zero angular momentum mode.
So long as the field theory interactions are generally covariant and
the background is fixed, the de Sitter boost isometry will be present.
It remains, then,
to find examples of interacting theories for which the zero
angular momentum sector has a conserved quantity. 
The explicit construction of such an interacting theory is an open
challenge.

\ack

IAM thanks the Kavli Institute for Theoretical Physics for its
hospitality during early stages of this project.
GC was supported by the West Chester University Summer Undergraduate
Research Institute (Summer 2019).

\section*{References}

\addcontentsline{toc}{section}{References}
\bibliographystyle{iopart-num}
\bibliography{myBib}

%
\end{document}